\documentclass[journal]{IEEEtran}
\IEEEoverridecommandlockouts

\usepackage{textcase}
\usepackage{changes}
\usepackage{gensymb}
\usepackage{cite}
\usepackage{amsmath,amssymb,amsfonts}
\usepackage{algorithmic}
\usepackage{textcomp}
\usepackage{graphicx}
\usepackage{amssymb}
\usepackage{amsfonts}
\usepackage{amsmath}
\usepackage{epsfig}
\usepackage{color}
\usepackage{fancybox}
\usepackage{textcomp}
\usepackage{multirow}
\usepackage{setspa ce}
\usepackage{psfrag}
\usepackage{booktabs}
\usepackage{float}
\usepackage[caption = false]{subfig}
\usepackage[ruled]{algorithm2e}

\SetKwInput{KwData}{\textbf{Initialization}}
\usepackage{mathrsfs}

    \def\Complex{{\rm\rule[.23ex]{.03em}{1.1ex}\kern-.3em{C}}}

    \newcommand{\be}{\begin{equation}} \newcommand{\ee}{\end{equation}}
    \newcommand{\bea}{\begin{eqnarray}} \newcommand{\eea}{\end{eqnarray}}
    \newcommand{\benum}{\begin{enumerate}} \newcommand{\eenum}{\end{enumerate}}

        \newcommand{\qc}{{\bf c}}

        \newcommand{\qn}{{\bf n}}

        \newcommand{\qv}{{\bf v}}
        
        \newcommand{\qx}{{\bf x}}
        \newcommand{\qy}{{\bf y}}

        \newcommand{\qH}{{\bf H}}
        \newcommand{\qI}{{\bf I}}

        \newcommand{\qQ}{{\bf Q}}

        \newcommand{\qV}{{\bf V}}

        \newcommand{\qSigma}{{\boldsymbol \Sigma}}

        \newcommand{\qOmega}{{\boldsymbol \Omega}}

        \newcommand{\qlambda}{{\boldsymbol \lambda}}
        \newcommand{\qgamma}{{\boldsymbol \gamma}}

        \newcommand{\qmu}{{\boldsymbol \mu}}

        \newcommand{\calN}{{\mathcal N}}

        \newcommand{\Ex}{{\sf E}}

        \newcommand*{\argmin}{\operatornamewithlimits{argmin}\limits}

\def\BibTeX{{\rm B\kern-.05em{\sc i\kern-.025em b}\kern-.08em
    T\kern-.1667em\lower.7ex\hbox{E}\kern-.125emX}}
\begin{document}

\title{{A Linear Bayesian Learning Receiver Scheme for Massive MIMO Systems }
\thanks{An extended version of this paper including all proofs is in preparation \cite{B-PIC2020}.
 This research was supported by an Australian Research Council Discovery Early Career Research Award (DE150101704) and the Research Training Program Stipend  from The University of Sydney.}}

\author{\IEEEauthorblockN{Alva Kosasih\IEEEauthorrefmark{1}, Wibowo Hardjawana\IEEEauthorrefmark{1}, Branka Vucetic\IEEEauthorrefmark{1}, Chao-Kai Wen\IEEEauthorrefmark{2} }\\
\IEEEauthorblockA{\IEEEauthorrefmark{1}Centre of Excellence in Telecommunications, The University of Sydney, Sydney, Australia. \\ \IEEEauthorrefmark{2}Institute of Communications Engineering, National Sun Yat-sen University, Kaohsiung, Taiwan. \\ 
Email: alva.kosasih@sydney.edu.au,  wibowo.hardjawana@sydney.edu.au, branka.vucetic@sydney.edu.au, chaokai.wen@mail.nsysu.edu.tw. }}

\maketitle

\begin{abstract}
Much stringent reliability and processing latency requirements in ultra-reliable-low-latency-communication (URLLC) traffic make the design of linear massive multiple-input-multiple-output (M-MIMO) receivers becomes very challenging. Recently, Bayesian concept has been used to increase the detection reliability in minimum-mean-square-error (MMSE) linear receivers. However, the latency processing time is a major concern due to the exponential complexity of matrix inversion operations in MMSE  schemes. This paper proposes an iterative M-MIMO receiver that is developed by using a Bayesian concept and a parallel interference cancellation  (PIC) scheme, referred to as a linear Bayesian learning (LBL) receiver. PIC has a linear complexity as it uses a combination of maximum ratio combining (MRC) and decision statistic combining (DSC) schemes to avoid matrix inversion operations.  Simulation results show that the bit-error-rate (BER) and latency processing performances of the proposed receiver outperform the ones of MMSE and best Bayesian-based receivers by minimum $2$ dB and $19$ times for various M-MIMO system configurations.
\end{abstract}


\section{Introduction}
Massive multiple-input-multiple-output (M-MIMO) technology has been proposed to support ultra reliability and low latency (URLLC) data transmissions.  The reduction of minimum transmission time intervals (TTI) and bit-error-rate (before coding) requirements 
from $15$ ms and  $10^{-3}$ to $1$ ms and  $10^{-5}$ in 4G and 5G NR URLLC \cite{Overview.M.MIMO,MIMO.5G,C.Sexton2017,M-MIMO_URLLC_mm-wave,M-MIMO_URLLC_wireless-access,M-MIMO_URLLC_2018}, respectively, pose a challenging problem in developing an M-MIMO receiver.
Two types of M-MIMO receivers have been investigated in the literature; classical and Bayesian learning receivers.

Classical receivers can be divided into two categories, non-linear and linear receivers. While non-linear receivers e.g. maximum likelihood (ML) receiver \cite{ML} can achieve an optimal symbol detection reliability,  it suffers a high computational complexity due to an exhaustive search operation to find the ML combinations of user symbols. This leads to a long detection processing time and thus a high latency. The second type is the linear M-MIMO receivers; 1) minimum mean square error (MMSE) and zero forcing (ZF) receivers \cite{LMMSE};  and 2) iterative parallel interference cancellation (PIC)  receivers that use a combination of maximum ratio combining (MRC) and decision statistic combining (DSC) schemes \cite{PIC2012}. MMSE and ZF receivers both rely on a matrix inversion operation to cancel multiple user interference which leads to an exponential increase in computational complexity with the number of antennas. In contrast,  PIC receivers avoid the use of a matrix inversion operation. Specifically, PIC receivers iteratively  performs  MRC to estimate multiple user symbols used to reconstruct interfering symbols. The interfering symbols are then subtracted from the received signal to recover the desired symbols, in a parallel manner. Consequently, PIC receivers have a linear computational complexity and thus a much lower latency. Note that all above linear receivers exhibit poor detection reliability when compared to an  ML receiver.

Recently, Bayesian learning concept has been introduced to reduce the reliability performance gap between ML and linear receivers. These receivers are referred to as the Bayesian learning receivers \cite{Jespedes-TCOM14, EP-CG2017, EPLowComplex2018,A.Kosasih,LAMA_Paper,OAMP_new}. The Bayesian learning concept is used to incorporate detection probability measures when estimating the detected symbols from the received signals\cite{Minka-01}. The best Bayesian learning receivers in term of bit-error-rate (BER) performance combines the Bayesian learning concept with MMSE scheme, referred to as expectation propagation (EP) receivers \cite{Jespedes-TCOM14, EP-CG2017, EPLowComplex2018,A.Kosasih}. Despite a significant performance improvement compared to the MMSE receivers, matrix inversion operations are still required. This results in a exponentially latency processing at the receivers. To avoid the matrix inversion operation, another Bayesian  based receiver referred to as approximate message passing (AMP) receiver \cite{LAMA_Paper,OAMP_new} has been proposed, albeit at the cost of higher BER as compared to EP receivers. Note also that all Bayesian M-MIMO receivers above rely on the learning parameters that needs to be searched by using a trial and error process for different environments.

We propose a novel iterative M-MIMO receiver referred to as linear Bayesian learning (LBL) receiver to cater for higher reliability and lower latency requirements in URLLC traffic. We first build the system model wherein an M-MIMO receiver at the base station is used to detect the symbols sent by multiple users. We assume that the channel estimates for different users are available and have been calculated by using different processes at the base stations. The developed LBL receiver consists of three modules; Bayesian symbol observation (BSO), Bayesian symbol estimate (BSE) and DSC modules. In the BSO module, we apply the MRC scheme to the received signals in order to get the symbol estimates, referred to as the observed symbols. For each observed symbol, PIC scheme is then used to remove its interference. The symbols variance is also calculated.
In the BSE module, the observed symbols and their variances are then utilised to construct maximum likelihood Gaussian distribution functions. The soft symbol estimates are then calculated based on the likelihood functions.  These estimates are used to compute symbol error between estimations and observations. In the DSC module, the value of symbol errors in current and previous iterations are used  to calculate the symbol estimates.  
 The process is then repeated iteratively and finally the DSC outputs are taken as the symbol estimates. The simulation results show that the
BER and latency performances of the proposed LBL receiver outperform existing classical and Bayesian M-MIMO receivers in \cite{LMMSE,PIC2012,LAMA_Paper} by minimum $2$ dB and $19 $ times for various M-MIMO system configurations.
The main contributions of this paper are as follows
\begin{itemize}
\item First Bayesian M-MIMO iterative receiver that uses PIC scheme \cite{PIC2012}. This leads to an elimination of the matrix inversion operations or approximations used in several advanced iterative receivers \cite{Jespedes-TCOM14,XYuan2014PA,EPLowComplex2018,A.Kosasih}. As a result, linearly computational latency processing and a near-optimal performance are achieved.
\item First Bayesian M-MIMO iterative receiver that derives the learning parameters directly from the symbol errors between estimations and observations in different iterations via the DSC scheme. This is in contrast to the trial and error process used by many Bayesian receivers \cite{Jespedes-TCOM14,EPLowComplex2018,OAMP_new,A.Kosasih}  to find the learning parameters.

\item We perform an analysis on the maximum number of users that can be supported by the proposed receiver in exchange of a low complexity signal processing. This analysis has not been discussed in the literature. 
\end{itemize}

{\bf Notations}---$\mathbf{I}$ denotes a proper size identity matrix.  For any matrix $\mathbf{A}$, $\mathbf{A}^{T}$ is the transpose of $\mathbf{A}$, $\mathbf{A}^{H}$ is the conjugate transpose of $\mathbf{A}$, and ${\sf tr}(\mathbf{A})$ denotes the traces of $\mathbf{A}$. 
${\rm diag}(\qQ)$ refers to the operation to force the non-diagonal elements of matrix $\qQ$ to zero.  
Conversely, ${\rm off}(\qQ)$ refers to the operation to force the diagonal elements of matrix $\qQ$ to zero.
 ${\Ex}[\qx]$ is the mean of random vector $\qx$ and ${\sf  Var}[\qx] = {\Ex}\big[\left(\qx-{\Ex}[\qx]\right)^2\big]$ is its variance.
  $\calN(x_k,c_k;v_k)$ represents a complex single variate Gaussian distribution  of random variable $x_k$ with mean $c_k$ and variance $v_k$. 
 By letting $\qx = [x_1, \cdots, x_K]^T, \qc = [c_1, \cdots, c_K]^T$, the multivariate Gaussian distribution of random vector $\qx$ is denoted as 
  \begin{equation}\label{Gaussian_Multi}
 \calN(\qx,\qc;\qSigma) = \frac{1}{\sqrt{(2\pi)^K |\qSigma|}} {\rm exp} \bigg(-\frac{1}{2} {\left({\qx-\qc}\right)^T \qSigma^{-1} \left({\qx-\qc}\right)} \bigg)
 \end{equation}
   $\qSigma$ is a covariance matrix, and $|\qSigma|$ is the determinant of the covariance matrix $\qSigma$. 
   
 \begin{figure}
\centering
{\includegraphics[scale=0.4]{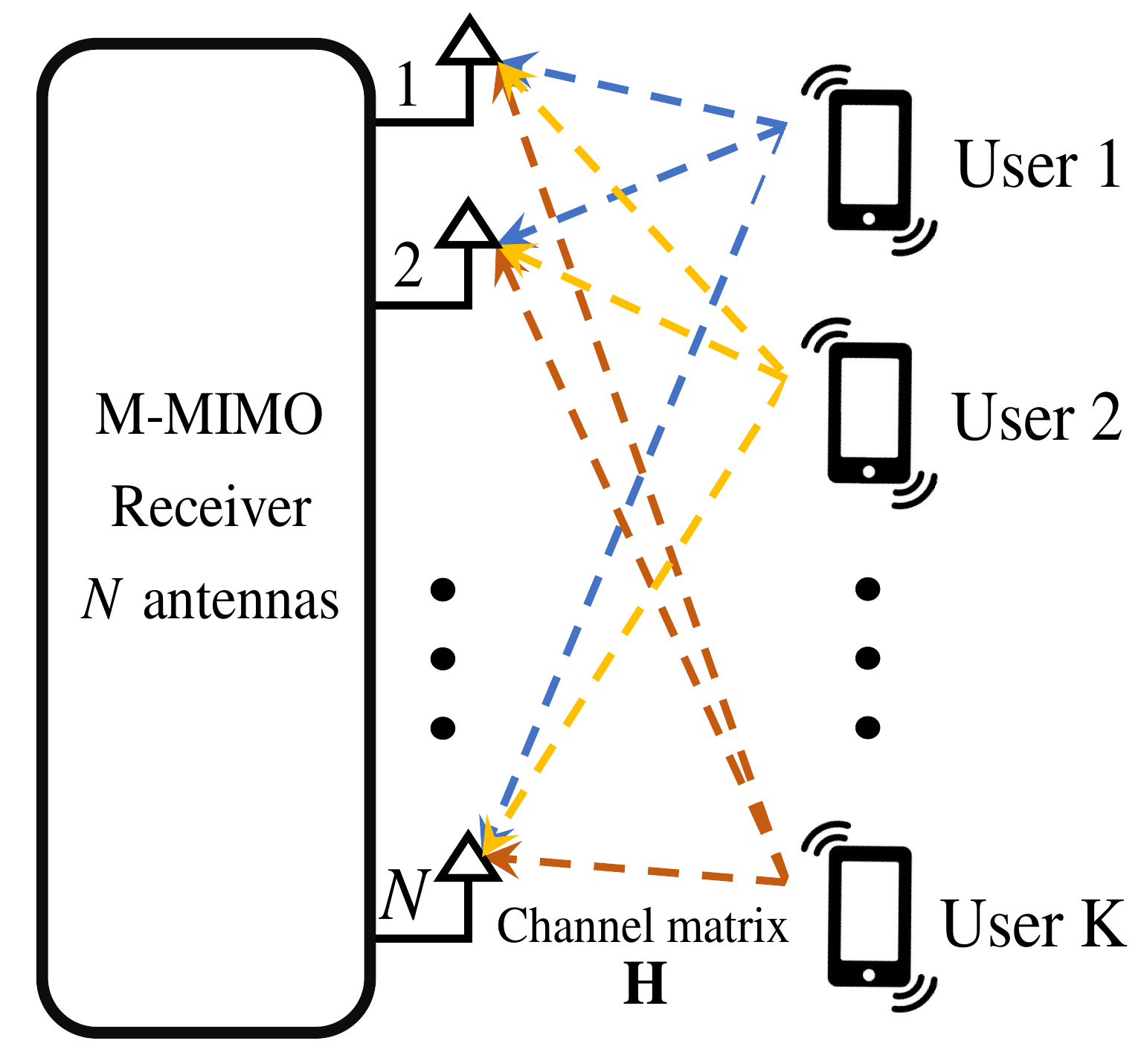}}\hfill
\caption{The uplink M-MIMO system.} 
\label{F00}
\end{figure} 

 \begin{figure}
\centering
{\includegraphics[scale=0.5]{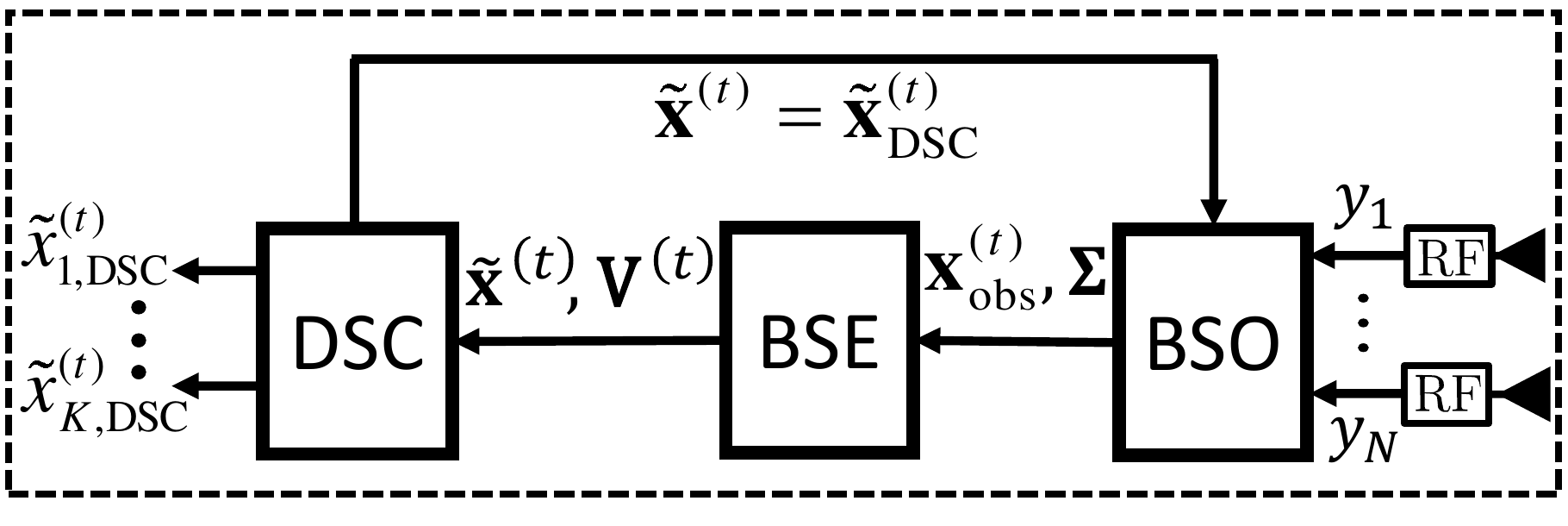}}\hfill
\caption{ System model of the LBL receiver. }
\label{F0}
\end{figure} 

\section{System Model}

We consider an  M-MIMO receiver that receives uplink signals from $K$ users, each with a single antenna, as depicted in  Fig. \ref{F00}. The receiver is equipped with a large number of antennas $N>>K$. Each user first maps its information bit stream to a symbol $x_k$ that belongs to a constellation point of $M$-QAM, $\Omega_k = [ s_1, \dots, s_M]$ where $\qx = [x_1, \cdots, x_K]^T$, $s_m$ is the $m$-th symbol in the $M$-QAM symbol constellations, and  $k=1,\dots,K$. The average symbol energy is $E_x = \Ex\{ |x_k|^2\}$. In the receiver side, the received signal at the M-MIMO receiver, $\bold{y}=[y_1 \ldots y_N]^{T}$ where $y_n$  is the received signal at antenna $n$ can then be written as
\begin{equation} \label{eII_1}
\qy = \qH \qx + \qn
\end{equation}
where, $\qH  \in \mathbb{C}^{N\times K} $ is the coefficients of complex Rayleigh wireless channels between $N$ antennas at M-MIMO receiver and $K$ users, $\qn \in \mathbb{C}^N$ denotes the additive white Gaussian noise (AWGN) with zero mean and covariance matrix $\sigma^2 \qI$.

The function used to cancel multiple user interference and  recover the symbols sent by $K$ users, $\bold{x}$ from the received signals, $\bold{y}$ when MMSE  receiver is used, is given as
\begin{equation}\label{LMMSE}
{\qx} \propto \left( \qH^H\qH +\sigma^{2} \mathbf{I} \right)^{-1}\qH^H \qy.
\end{equation}
Note that the matrix inversion operation used in \eqref{LMMSE} is costly as its complexity increases exponentially with the number of antennas.
 In contrast, when the iterative PIC receiver is used there is no matrix inversion operation performed. At the M-MIMO receiver, the symbols recovery from $ \bold{y}$  in iteration $t$, $\qx^{(t)}$ is given as 
\begin{equation}\label{PIC-EP}
\qx^{(t)}\propto{\rm diag}^{(-1)}(\qH^H\qH)(\qH^H\qy-{\rm off}(\qH^H\qH)\tilde{\qx}^{(t-1)} )
\end{equation} 
where $\tilde{\qx}^{(t-1)}  = [\tilde{x}_1^{(t-1)} , \cdots, \tilde{x}_{K}^{(t-1)} ]^T $ are the symbol estimates from previous iteration $t-1$.
Note that to estimate $\qx^{(t+1)}$, we set $\tilde{\qx}^{(t)}=\qx^{(t)}$. 

{\section{Expectation Propagation Receiver}

In this section, we briefly review the best Bayesian based receiver in the literature, Expectation Propagation (EP) receivers \cite{Jespedes-TCOM14, EP-CG2017, EPLowComplex2018,A.Kosasih,EP-NSA2018,EP2019}, which is a combination of Bayesian and MMSE concepts. 
The main idea of the EP is to iteratively approximate the distribution of a random transmitted symbol vector, $\qx$ by using a Gaussian probability distribution function (PDF) approximation based on the received signals $\qy$, $p^{(t)}(\qx|\qy)$,  and a pair of tuning parameters, $(\qlambda^{(t)},\qgamma^{(t)})$ which are obtained from the exponential family distributions \cite{Minka-01}. $  p^{(t)}(\qx|\qy)$ is given as
\begin{flalign} \label{eq:postPro}
    p^{(t)}(\qx|\qy)  = \mathcal{N} \left(\qx, {\qmu}^{(t)}_{\rm obs}; \qSigma^{(t)}_{\rm obs} \right)
\end{flalign}
with its mean and variance are obtained  from the received signal observation, $\qy$,
\begin{subequations} \label{eEP_0102}
            \begin{align}
&\qmu^{(t)}_{\rm obs}   = {\qSigma}^{(t)}_{\rm obs}   {\left(\sigma^{-2} \qH^H\mathbf{y} + \qgamma^{(t)}\right)} \label{eEP_a02}\\
& {\qSigma}^{(t)}_{\rm obs} =  {\left(\sigma^{-2} \qH^H\qH + \qlambda^{-1(t)}\right)}^{-1}. \label{eEP_a01}
            \end{align}
        \end{subequations}
We then construct a Gaussian likelihood function to approximate the distribution of received signal $\qy$ for a given statistic of $\qx$ in \eqref{eEP_0102}. This is  denoted as $ p^{(t)}(\qy|\qx)$,
 \begin{flalign} \label{eq:postPr}
    p^{(t)}(\qy|\qx) = \mathcal{N} \left( \qx,\qx_{\rm ext}^{(t)};\qv^{(t)}_{{\rm ext}}\right)  
\end{flalign}
where $\mathcal{N} \left( \qx,\qx_{\rm ext}^{(t)};\qv^{(t)}_{{\rm ext}}\right) = \prod_{k=1}^{K} \mathcal{N}\left(x_k,x_{k,{\rm ext}}^{(t)};v_{k,{\rm ext}}^{(t)}\right)$ and
\begin{subequations} \label{eEP_a0304}
            \begin{align}
&x_{k,{\rm ext}}^{(t)} = v_{k,{\rm ext}}^{(t)} {\left(\frac{\mu_{k,{\rm obs}}^{(t)}}{\Sigma_{k,{\rm obs}}^{(t)}}-\gamma^{(t)}_{k}\right)} \label{eEP_a04}\\
&v_{{k,{\rm ext}}}^{(t)} =  \frac{\Sigma_{k,{\rm obs}}^{(t)}}{1- \Sigma_{k,{\rm obs}}^{(t)} \lambda_{k}^{(t)}}.  \label{eEP_a03}
            \end{align}
        \end{subequations}
Here, $x_{k,{\rm ext}}^{(t)}$, $\mu^{(t)}_{k,{\rm obs}}$ and $\gamma^{(t)}_{k}$ are the $k-$th element of vectors $\qx_{{\rm ext}}^{(t)}$, $\qmu^{(t)}_{\rm obs}$ and $\qgamma^{(t)}$, respectively. $v_{{k,{\rm ext}}}^{(t)}, \Sigma^{(t)}_{k,{\rm obs}}$, and $\lambda_k^{(t)}$ are the $k$-th diagonal element of matrices $\qv_{{{\rm ext}}}^{(t)}, \qSigma^{(t)}_{\rm obs}$, and $\qlambda^{(t)}$, respectively. 

Now, we estimate the soft symbols of $\qx$ based on the Gaussian likelihood function in \eqref{eq:postPr}. The soft symbol estimates and its variance are given as  
\begin{subequations}\label{eEP_b0102}
            \begin{multline}\label{eEP_b01}
\hat{\qx}^{(t)}= \Ex {\left[ \qx | \qx_{\rm ext}^{(t)} ,\qv_{\rm ext}^{(t)}  \right]} = c  \sum_{\qx \in \qOmega} \qx  \text{  } \mathcal{N} \left( \qx,\qx_{\rm ext}^{(t)};\qv^{(t)}_{{\rm ext}}\right)
            \end{multline}
            \begin{multline}\label{eEP_b02}
\qv^{(t)} = {\sf Var} {\left[ \qx | \qx_{\rm ext} ^{(t)}, \qv_{\rm ext}^{(t)} \right]} =    \\
\Ex \left[ \left| \qx  - \Ex  \left[\qx | \qx_{\rm ext}^{(t)},\qv_{\rm ext}^{(t)} \right] \right|^{2} \right],
            \end{multline}
\end {subequations}
respectively, where $c$ is a normalisation constant to ensure that the summation of $  p^{(t)}(\qy|\qx) $ is $1$. 

The iteration of the EP receiver is performed until the values of $\hat{\qx}^{(t)}$ and $\qv^{(t)} $ in \eqref{eEP_b0102} are close to those of $\qmu^{(t)}_{\rm obs}$ and $\qSigma^{(t)}_{\rm obs}$ in \eqref{eEP_0102}, respectively. This is referred to  as the moment matching condition \cite{Jespedes-TCOM14}. When the moment matching condition has not been satisfied, the EP receiver recomputes the parameters $\left(\qgamma^{(t+1)},\qlambda^{(t+1)} \right)$ in \eqref{eEP_0102} by using \eqref{eEP_b0102} and a predefined learning parameter, $\beta$. Note that $\beta$ is used to weight the  parameters,  $\left(\qgamma^{(t+1)},\qlambda^{(t+1)}\right)$ in two consecutive iterations. 

Their details are described in \cite{Jespedes-TCOM14}. Although EP receivers can achieve near optimal detection performance \cite{Jespedes-TCOM14, EP-CG2017, EPLowComplex2018,A.Kosasih,EP-NSA2018,EP2019}, it suffers a highly computational complexity as it performs the matrix inversion operation \eqref{eEP_a01}, in every iteration.  Furthermore, the predefined learning parameter  needs to be searched beforehand by using the trial and error processes. These limitations prohibit the deployment of EP receivers in real time systems.

\section{Linear Bayesian Learning M-MIMO Receiver}

In this section, we propose a novel linear Bayesian based receiver referred to as a linear Bayesian learning (LBL) receiver which avoids the matrix inversion operations in \eqref{eA1_a01} and the search of a learning parameter. The proposed receiver  is shown in Fig. \ref{F0}. It consists of three modules, BSO module that computes the probability distribution function (PDF) of observed symbols from the  received signals; BSE module that yields soft symbol estimates based on the computed PDFs; DSC module refines the symbol estimates by using the BSE outputs and returns the refined estimates to BSO module.
 
\subsection{Bayesian Symbol Observation}

The computation of PDF of symbols from the observed received signal is done by treating  $\qx$ in \eqref{eII_1} at each iteration $t$ as a random vector. Its mean and variance are obtained from observation $\qy$ based on \eqref{PIC-EP}. The PDF of symbol vector $\qx$ for a given $\qy$ at iteration $t$, $p^{(t)}(\qx|\qy)$ is given as
\begin{equation}\label{dist_of_x_y}
 p^{(t)}({\qx}|\qy) = \mathcal{N}\big(\qx,\qx_{\rm obs}^{(t)}; \qSigma)
\end{equation}
where
\begin{subequations} \label{eA1_a0102}
            \begin{align}
&\qSigma = {\sf  Var}[\qx] \propto \left({\sigma^{-2} {\rm diag}(\qH^H\qH)}\right)^{-1},  \label{eA1_a01}\\
&\qx_{\rm obs}^{(t)}= {\Ex}[\qx]   =\qSigma \sigma^{-2} { \left( \qH^H\mathbf{y}-{\rm off }(\qH^H\qH) \tilde{\qx}^{(t-1)} \right)} \label{eA1_a02}
            \end{align}
        \end{subequations}
 where   $\Sigma_k$ is the $k$-th diagonal element of matrix $\qSigma$, $x_{{\rm obs},k}^{(t)}$ is the $k$-th element of vector $\qx_{\rm obs}^{(t)}$. Note that $\qSigma$  in \eqref{eA1_a01} is a diagonal matrix and thus no matrix inversion operation is needed. This is in contrast to the matrix inversion operations in EP receiver \eqref{eEP_a01}. The results, $\left(\qx_{\rm obs}^{(t)},\qSigma\right)$ are then forwarded to BSE module, as shown in Fig. \ref{F0}.
  
  \subsection{Bayesian Symbol Estimator}
      
In the BSE module, in Fig. \ref{F0}, we compute the soft symbol estimate, $\tilde{\qx}^{(t)}$. We use \eqref{dist_of_x_y} and \eqref{eA1_a0102}  obtained from the BSO module to write the likelihood probability function of symbol  $x_k$ of user $ k $,
\begin{equation}\label{eI_53} \notag
p(x_k|x_{{\rm obs},k}^{(t)},\Sigma_k) = c  \times \calN{\left( x_k; x_{{\rm obs},k}^{(t)}, \Sigma_k\right)}
\end{equation}
where  $c$ is a normalisation constant to ensure that $ \sum_{x_k \in \Omega_k} p(x_k|x_{{\rm obs},k}^{(t)},\Sigma_k) = 1$.
The soft symbol estimate, $\tilde{x}_k^{(t)}$ is then given as 
            \begin{multline}\label{eA1_b01}
\tilde{x}_k^{(t)}= \Ex {\left[ x_k |x_{{\rm obs},k}^{(t)} ,\Sigma_k \right]} = \sum_{x_k \in \Omega_k}  x_k {p}{\left(x_k|x_{{\rm obs},k}^{(t)},\Sigma_k\right)},
            \end{multline}
  $\tilde{\qx}^{(t)} = [\tilde{x}_1^{(t)}, \dots, \tilde{x}_K^{(t)}]$ and $\Omega_k$ is the $k$-th constellations. The errors between estimations and observations for $K$ users  at iteration $t$, $\qV^{(t)} $ is given as
            \begin{multline}\label{DSC_Var}
\qV^{(t)} = \\ {\rm diag}
\left(\left(\qH^H\qy - (\qH^H\qH) \tilde{\qx}^{(t)}\right)
\left(\qH^H\qy - (\qH^H\qH) \tilde{\qx}^{(t)}\right)^T\right).
            \end{multline} 
 The outputs of BSE module $\left(\tilde{\qx}^{(t)},\qV^{(t)}\right)$ from \eqref{eA1_b01} and \eqref{DSC_Var} are sent to the DSC  module.

\subsection{Decision Statistic Combining}

 In this section, we use the  DSC scheme to approximate the learning parameter based on the values of instantaneous errors in subsequent iterations. This is shown in Fig. \ref{F0} and done  by weighting the outputs of the BSE module, $\tilde{\qx}$ in the current and previous iterations based on the error $\qV$ in  \eqref{DSC_Var}, given as
\begin{multline}\label{DSC}
                   \qx_{\rm DSC}^{(t)} = \left( \qV^{(t)}+\qV^{(t-1)}\right)^{-1}   \left(   \qV^{(t-1)}  \tilde{\qx}^{(t)}  +       \qV^{(t)}  \tilde{\qx}^{(t-1)} \right)
\end{multline}
where $ {\qx}_{\rm DSC}^{(t)} = [ x_{1,{\rm DSC}}^{(t)}, \dots, x_{K,{\rm DSC}}^{(t)}]$.
The iterative process will stop if the following is achieved,
\begin{equation}\label{eq_convergence}
 \|\qx_{\rm DSC}^{(t)} - \qx_{\rm DSC}^{(t-1)} \| \leq \epsilon  \lor t = T_{\rm max}
\end{equation}
 where  $\epsilon$ is the minimum acceptable difference of the $\qx^{(t)}_{\rm DSC}$ in two consecutive iterations and $T_{\rm max}$ is the maximum number of the iterations.
 If \eqref{eq_convergence} is not satisfied, we use $\qx_{\rm DSC}^{(t)} $  as the inputs of BSO in the next iteration
\begin{equation}\label{Assign}
\tilde{\qx}^{(t)} =  {\qx}_{\rm DSC}^{(t)},
\end{equation}
 else the symbol hard decision is then made by using the outputs of DSC in \eqref{DSC},
 \begin{equation}\label{Hard_Dec}
 \qx_{\rm hard} =\argmin \| \qx_{{\rm DSC}}^{(T)} - \qOmega  \|^2
 \end{equation}
 where $T$ is the last number of iteration and $\qOmega$ is the set of the constellations. Note that the use of \eqref{DSC_Var} and  \eqref{DSC} eliminates the trial and error issues for finding optimal learning parameters used in most of Bayesian learning iterative receivers, Eq.  (37) and (38) in \cite{Jespedes-TCOM14}, Eqs.  (15) and (16) in\cite{EPLowComplex2018}, and  Eq. (8) in \cite{OAMP_new}.  
The complete pseudo-code is shown in Alg. \ref{A1}.
\begin{algorithm}
    \caption{LBL Algorithm \label{A1}}
    \small
    \KwData{ $\tilde{\qx}^{(0)}=0,T_{\rm max}=10$\;}
   \While{{\rm \eqref{eq_convergence} is not satisfied}}{
        1. Compute the mean and variance, $\qmu^{(t)},\qSigma$, in \eqref{eA1_a0102}
                
        2. Compute  the mean and variance, $\hat{\qx}^{(t)},\hat{\qV}^{(t)}$, in \eqref{eA1_b01},\eqref{DSC_Var}
        
        3. Compute $\qx_{\rm DSC}^{(t)} $, in \eqref{DSC}  
        
        4. Execute \eqref{Assign}
            } 
       \text{ }  5. Execute \eqref{Hard_Dec}				 \\     
\end{algorithm}

\section{Complexity and Convergence Analysis}

\begin{figure}
\centering
{\includegraphics[scale=0.54]{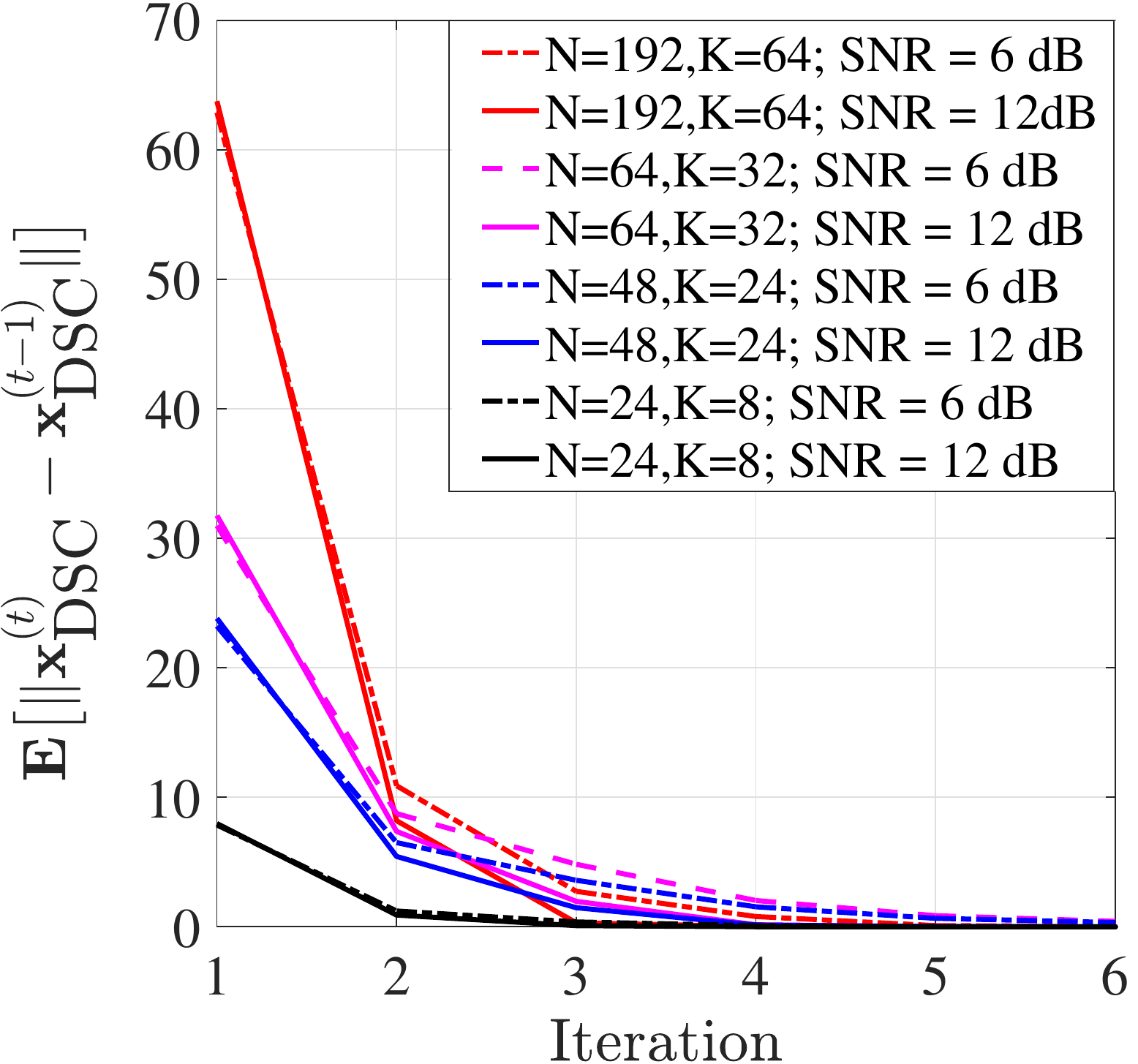}}\hfill
  \caption{The Convergence behaviour of the LBL receiver. }
  \label{F5}
\end{figure}

\begin{table}\large
\centering
  \caption{Computational complexity comparison.}
  \label{table1}
    \begin{tabular}{|l|c|r|}
        		\hline 
 Receiver 																					&     	Complexity		\\ 
        		\hline 
   		\hline 
    		LBL 																					& 		$\mathcal{O} (NKT)$ 				\\
     	 \hline
    		AMP \cite{LAMA_Paper}														& 		$\mathcal{O} (NKT)$ 	 			\\
    		\hline
     	MMSE 																				& 		$\mathcal{O} (N^2K+NK)$   			\\
     	 \hline
     	EP in  \cite{Jespedes-TCOM14} 											& 		$\mathcal{O} ((N^2K+NK)T)$ 	 		\\
     	 \hline
     	EP in   \cite{A.Kosasih} 														& 		$\mathcal{O} ((NK^2+NK)T)$ 	 		\\
         \hline
     	ML \cite{ML}																		& 		$\mathcal{O} (M^K)$  		\\ 
     	\hline
    		\hline
  \end{tabular}
\end{table}

We first evaluate the computational complexity of the LBL receiver.The computational complexity of the proposed receiver; the representative of best linear receivers, MMSE  \cite{LMMSE} scheme; the representative of  Bayesian receivers EP \cite{Jespedes-TCOM14,A.Kosasih} and AMP \cite{LAMA_Paper} schemes; and the exhaustive search based ML scheme \cite{ML} are tabulated in Table I. The table indicates that the computational complexity of the proposed receiver increases linearly with the number of antennas, $N$ and users, $K$ by avoiding matrix inversion operations. This is in contrast to all other receivers whose computational complexity grow  exponentially with $N$ and/or $K$. We have also observed that the number of iterations, $T$ needed for the EP, AMP, and LBL receivers  are similar. However, the convergence rates of the  AMP  and EP  receivers  are not shown in this paper due to a space limitation. Therefore, the proposed receiver has a significantly lower processing latency and thus more suitable for URLLC data traffic. For instance, for $K=64, N=192, \text{and } T=10$, LBL receiver complexity is 19 times lower than MMSE receiver, and 192 times lower than EP receiver. We conclude that by eliminating the matrix inversion operations, we can obtain a significantly lower computational complexity and thus much lower latency. 

To analyse the convergence rate of the proposed LBL receiver, we plot the value of the difference between symbol estimates, $\qx_{{\rm DSC}}$ in iteration $t$ and $t-1$ versus the iteration $t$ for $100,000$ channel realisations in Fig. \ref{F5}, and  $\epsilon = 10^{-6}$. The figure shows that the maximum number of iterations needed is  $6$ for various system configurations, implying a low latency. This also indicates that the iterations needed is relatively insensitive to the system configurations.

\section{Reliability Analysis}

To analyse receivers’ reliability, we plot the BER performances of our proposed receiver versus the linear receivers based on PIC  \cite{PIC2012}  and MMSE \cite{LMMSE} schemes; Bayesian learning receivers based on EP \cite{Jespedes-TCOM14,A.Kosasih} and AMP\cite{LAMA_Paper}  schemes;  an optimal ML receiver \cite{ML} for different system configurations. The same modulation scheme and number of channel realisation as in Section V are used. Note also that we perform trial and error process to search for the optimal learning parameters for EP and AMP off-line. We found that the optimal learning parameters in the EP and AMP receivers are $\beta = 0.9 $ and $0.2$, respectively. 

In Fig. \ref{F4}a, we employ the number of receive antennas, $N=24$ and users, $ K=8$. It shows that the LBL receiver outperforms (MMSE and PIC) receivers for about 2 dB to achieve BER $10^{-4}$. In practice, this also implies that the radio coverage of the proposed M-MIMO receiver is $26\%$ broader than those of the existing linear receivers.  Note also the similarity in term of BER performance with the EP and ML receivers that have  significantly higher computational complexity and latency as explained in the previous section. In Fig. \ref{F4}b, the number of received antennas and users are increased to $N=192$ and $K=64$. Similar behaviour is observed in relation to the BER performance in comparison with the linear and Bayesian learning receivers. Note that in Fig. \ref{F4}b we are unable to draw ML receivers due to its high computational complexity. 
From these facts, we conclude that the LBL receiver is more reliable than several existing classical and advance iterative receivers.

In Fig. \ref{F6}, we analyse the relationship between the number of users and used received antennas in the proposed M-MIMO receiver. We first define the ratio between the number of users and used received antennas as $\alpha$ in Fig. \ref{F6}. In Fig. \ref{F6}a, we set a fixed number of antennas, $N=256$. The EP and LBL receivers can achieve BER less than $10^{-4}$ when $\alpha <0.71$ and $\alpha <0.65$, respectively. In Fig. \ref{F6}b, we increase the number of antennas, $N=1024$ and obtain $\alpha <0.61$ and $0.7$ for LBL and EP receivers, respectively.  This implies that the exchange of the low complexity signal processing of the proposed receiver is the number of supported users.

\begin{figure}
\centering
\subfloat[ $N=24, K=8$, $M=4.$]
{\includegraphics[scale=0.48]{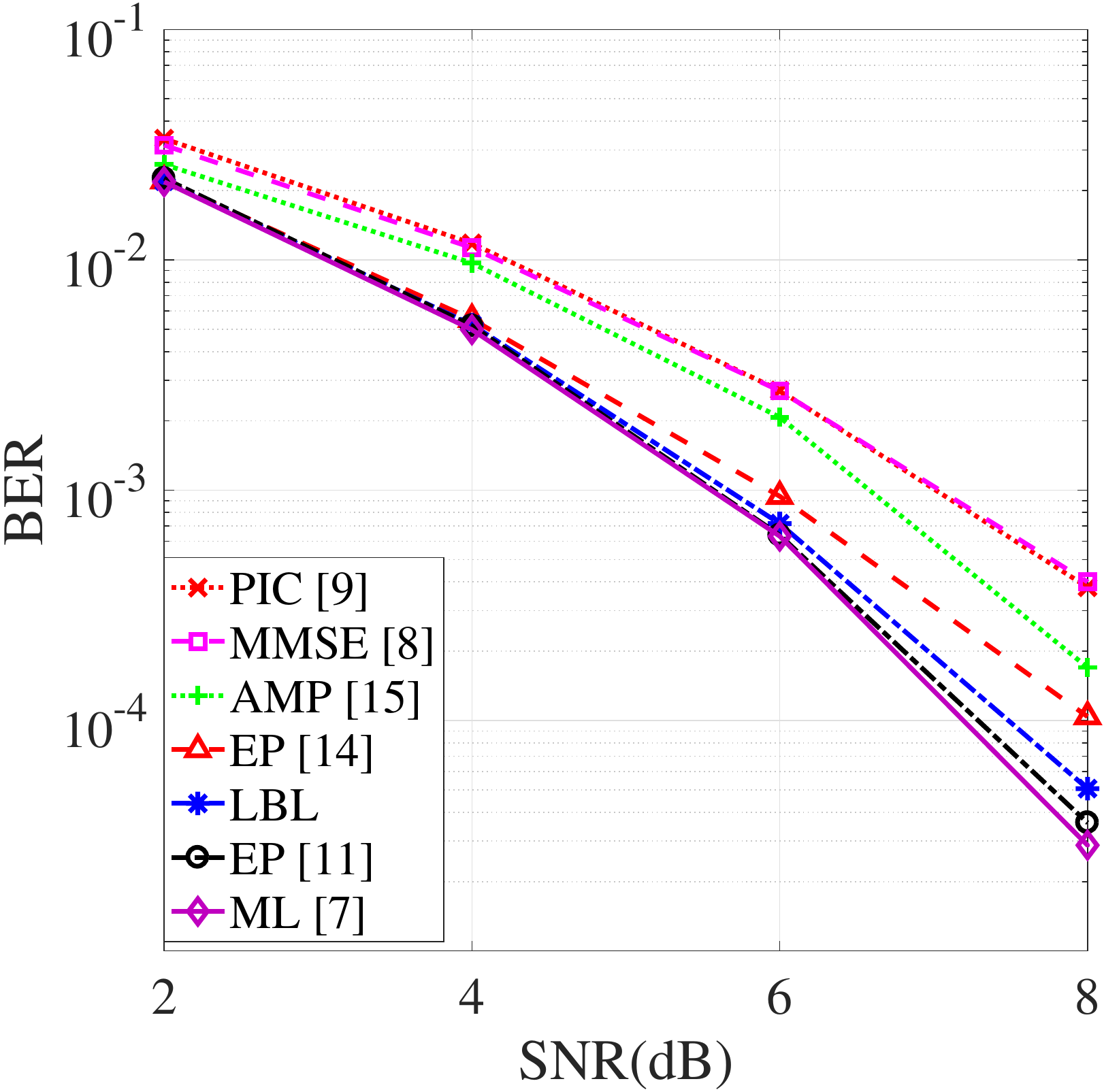}}\hfill
\centering
\subfloat[$N=192, K=64$, $M=4.$]
{\includegraphics[scale=0.48]{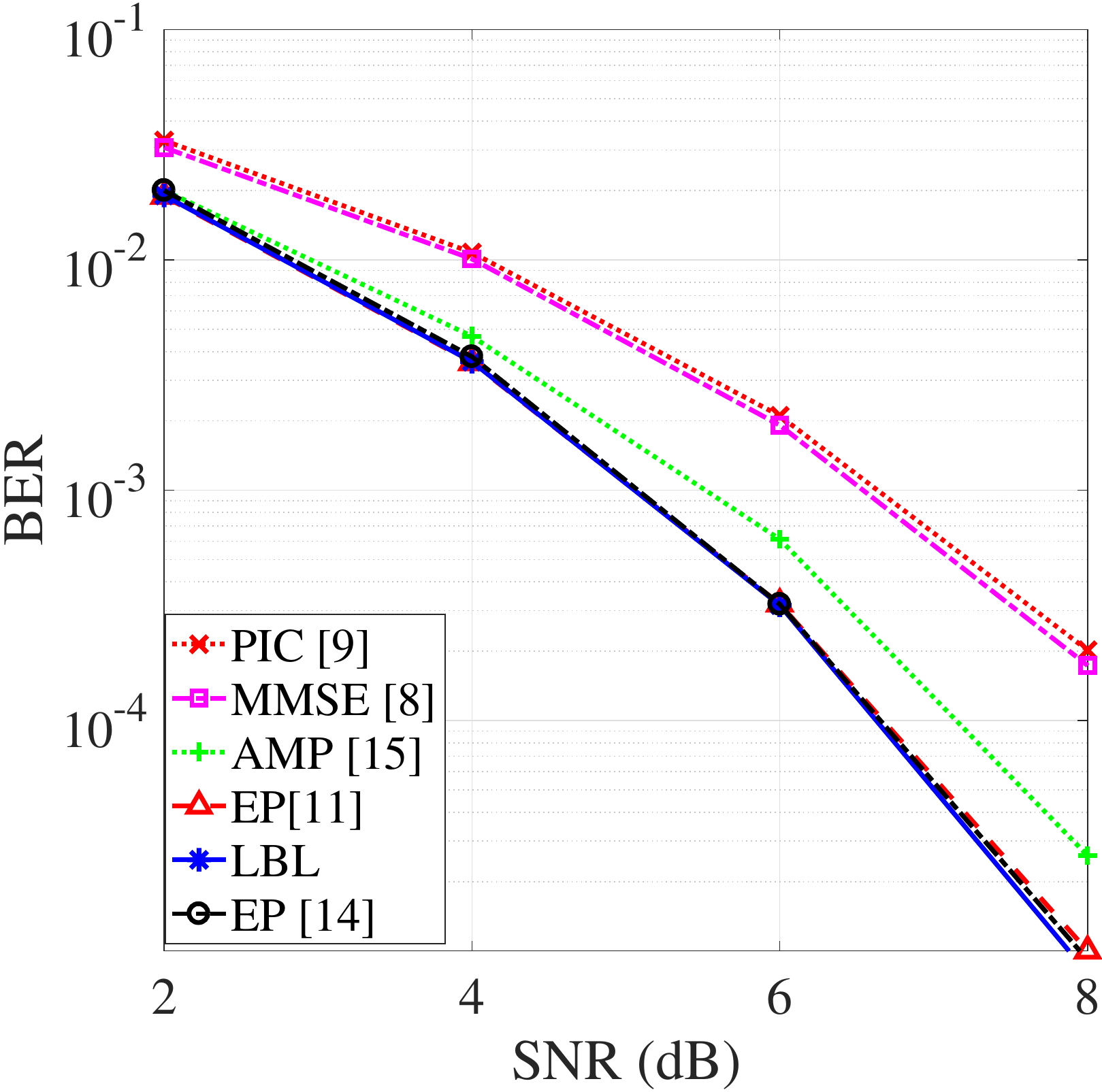}}
\caption{The BER performance of PIC, MMSE, AMP,  EP, LBL, ML receivers in several system configurations.}
\label{F4}
\end{figure}

\begin{figure}
\centering
\subfloat[ $N=256$ and $M=4.$]
{\includegraphics[scale=0.433]{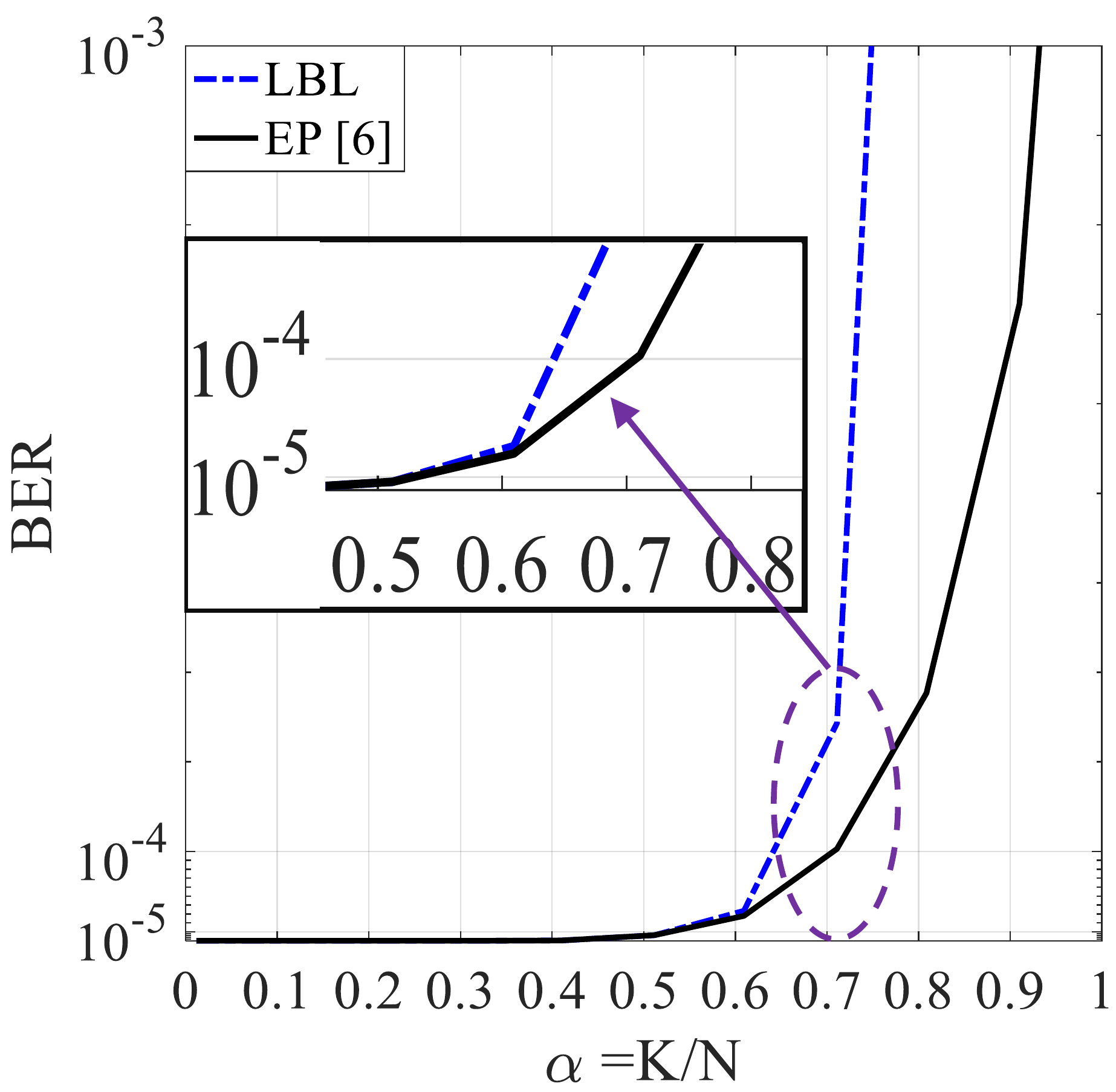}}\hfill
\centering
\subfloat[$N=1024$ and $M=4.$]
{\includegraphics[scale=0.433]{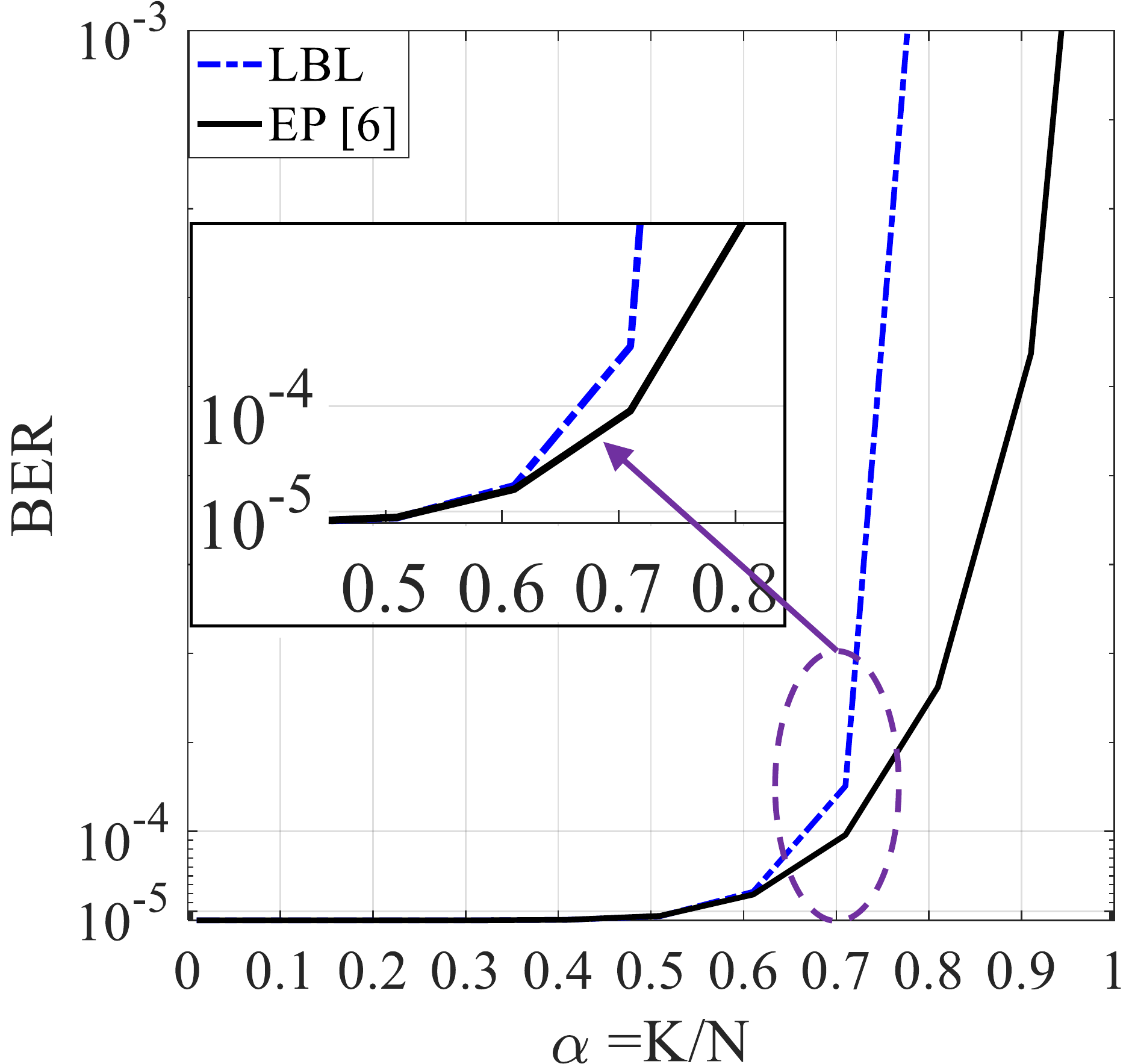}}
\caption{The BER performance of LBL and EP receivers with respect to the ratio of users to antennas, $\alpha$.}
\label{F6}
\end{figure}

\section{Conclusion}

We propose a novel Bayesian-based parallel interference cancellation receiver referred to as the LBL receiver. Simulation results show that the BER performance of the LBL receiver is very close to that of maximum likelihood receiver while maintaining a linear latency processing time in contrast to other existing schemes in the literature.

{\renewcommand{\baselinestretch}{1.1}
\begin{footnotesize}
\bibliographystyle{IEEEtran}
\bibliography{IEEEabrv,myBib}

% Generated by IEEEtran.bst, version: 1.14 (2015/08/26)
\begin{thebibliography}{10}
\providecommand{\url}[1]{#1}
\csname url@samestyle\endcsname
\providecommand{\newblock}{\relax}
\providecommand{\bibinfo}[2]{#2}
\providecommand{\BIBentrySTDinterwordspacing}{\spaceskip=0pt\relax}
\providecommand{\BIBentryALTinterwordstretchfactor}{4}
\providecommand{\BIBentryALTinterwordspacing}{\spaceskip=\fontdimen2\font plus
\BIBentryALTinterwordstretchfactor\fontdimen3\font minus
  \fontdimen4\font\relax}
\providecommand{\BIBforeignlanguage}[2]{{%
\expandafter\ifx\csname l@#1\endcsname\relax
\typeout{** WARNING: IEEEtran.bst: No hyphenation pattern has been}%
\typeout{** loaded for the language `#1'. Using the pattern for}%
\typeout{** the default language instead.}%
\else
\language=\csname l@#1\endcsname
\fi
#2}}
\providecommand{\BIBdecl}{\relax}
\BIBdecl

\bibitem{B-PIC2020}
{A. Kosasih, W. Hardjawana, C. K. Wen, and B.Vucetic}, ``{Bayesian-based
  parallel interference cancellation receiver for massive MIMO systems},''
  \emph{in preparation for a journal}.

\bibitem{Overview.M.MIMO}
{L. Lu, G. Y. Li, A. L. Swindlehurst, A. Ashikhmin, and R. Zhang}, ``{An
  overview of massive MIMO: Benefits and challenges},'' \emph{IEEE J. Sel.
  Topics Signal Process.}, vol.~8, no.~5, pp. 742--758, Oct. 2014.

\bibitem{MIMO.5G}
{Y. Huo, X. Dong, and W. Xu}, ``{5G cellular user equipment: From theory to
  practical hardware design},'' \emph{IEEE Access}, vol.~5, pp.
  13\,992--14\,010, Jul. 2017.

\bibitem{C.Sexton2017}
{C. Sexton, N. J. Kaminski, J. M. Marquez-Barja, N. Marchetti, and L. A.
  DaSilva}, ``{5G: Adaptable networks enabled by versatile radio access
  technologies},'' \emph{IEEE Commun. Surveys Tuts.}, vol.~19, no.~2, pp.
  688--720, 2nd Quart. 2017.

\bibitem{M-MIMO_URLLC_mm-wave}
{T. K. Vu, C. Liu, M. Bennis, M. Debbah, M. Latva-aho, and C. S. Hong},
  ``{Ultra-reliable and low latency communication in mmWave-enabled massive
  MIMO networks},'' \emph{IEEE Commun. Lett.}, vol.~21, no.~9, pp. 2041--2044,
  Sep. 2017.

\bibitem{M-MIMO_URLLC_wireless-access}
{P. Popovski, Č. Stefanović, J. J. Nielsen, E. de Carvalho, M.
  Angjelichinoski, K. F. Trillingsgaard, and A. Bana}, ``{Wireless access in
  ultra-reliable low-latency communication (URLLC)},'' \emph{IEEE Trans.
  Commun.}, vol.~67, no.~8, pp. 5783--5801, Aug. 2019.

\bibitem{M-MIMO_URLLC_2018}
{A. Bana, G. Xu, E. D. Carvalho, and P. Popovski}, ``{Ultra reliable low
  latency communications in massive multi-antenna systems},'' in \emph{Proc.
  52nd Asilomar Conf. Signals, Syst. and Comput.}, USA, Oct. 2018, pp.
  188--192.

\bibitem{ML}
{U. Fincke and M. Pohst}, ``{Improved methods for calculating vectors of short
  length in a lattice, including a complexity analysis},'' \emph{Math.
  Comput.}, vol.~44, no. 170, p. 463–471, Apr. 1985.

\bibitem{LMMSE}
{G. Caire, R. Muller, and T. Tanaka}, ``{Iterative multiuser joint decoding:
  Optimal power allocation and low-complexity implementation},'' \emph{IEEE
  Trans. Inf. Theory}, vol.~50, no.~9, p. 1950–1973, Sep. 2004.

\bibitem{PIC2012}
{N. Aboutorab, W. Hardjawana, and B. Vucetic}, ``{A new iterative
  doppler-assisted channel estimation joint with parallel ICI cancellation for
  high-mobility MIMO-OFDM systems},'' \emph{IEEE Trans. Veh. Technol.},
  vol.~61, no.~4, pp. 1577--1589, May 2012.

\bibitem{Jespedes-TCOM14}
{J.~C\'{e}spedes, P. M.~Olmos, M.~S\'{a}nchez-Fern\'{a}ndez, and
  F.~P\'{e}rez-Cruz}, ``{Expectation propagation detection for high-order
  high-dimensional MIMO systems},'' \emph{IEEE Trans. Commun.}, vol.~62, no.~8,
  pp. 2840--2849, Aug. 2014.

\bibitem{EP-CG2017}
{K. Takeuchi and C. K. Wen}, ``{Rigorous dynamics of expectation-propagation
  signal detection via the conjugate gradient method},'' in \emph{Proc. IEEE
  18th Int. Workshop Signal Process. Advances in Wireless Commun.}, Japan, July
  2017, pp. 1--5.

\bibitem{EPLowComplex2018}
{G. Yao, G. Yang, J. Hu, and C. Fei}, ``{A low complexity expectation
  propagation detection for massive MIMO system},'' in \emph{Proc. IEEE Global
  Commun. Conf.}, UAE., Dec. 2018, pp. 1--6.

\bibitem{A.Kosasih}
{H. Wang, A. Kosasih, C. K. Wen, J. Shi, and W. Hardjawana}, ``{Expectation
  propagation detector for extra-large scale massive MIMO},'' \emph{submitted
  for publication}.

\bibitem{LAMA_Paper}
\BIBentryALTinterwordspacing
{C. Jeon, R. Ghods, A. Maleki, and C. Studer}. (2018) {Optimal data detection
  in large MIMO}. [Online]. Available: \url{https://arxiv.org/abs/1811.01917}
\BIBentrySTDinterwordspacing

\bibitem{OAMP_new}
{J. Zhang, C. K. Wen, S. Jin, and G. Y. Li}, ``{Artificial intelligence-aided
  receiver for a CP-free OFDM system: Design, simulation, and experimental
  test},'' \emph{IEEE Access}, vol.~7, pp. 58\,901--58\,914, May 2019.

\bibitem{Minka-01}
{Minka, Thomas P}, ``{Expectation propagation for approximate Bayesian
  inference},'' in \emph{Proc. 17th Conf. Uncertainty in Artificial Intell.},
  USA, Aug. 2001, pp. 362--369.

\bibitem{XYuan2014PA}
{X. Yuan, J. Ma, and L. Ping}, ``{Energy spreading transform based MIMO
  systems: Iterative equalization, evolution analysis, and precoder
  optimization},'' \emph{IEEE Trans. Wireless Commun.}, vol.~13, no.~9, pp.
  5237--5250, Sept. 2014.

\bibitem{EP-NSA2018}
{Y. Zhang, Z. Wu, C. Li, Z. Zhang, X. You, and C. Zhang}, ``{Expectation
  propagation detection with neumann-series approximation for massive MIMO},''
  in \emph{Proc. IEEE Int. Workshop Signal Process. Syst.}, South Afr., Oct.
  2018, pp. 59--64.

\bibitem{EP2019}
{X. Tan, Y. Ueng, Z. Zhang, X. You, and C. Zhang}, ``{A low-complexity massive
  MIMO detection based on approximate expectation propagation},'' \emph{IEEE
  Trans. Veh. Technol.}, vol.~68, no.~8, pp. 7260--7272, Aug. 2019.

\end{thebibliography}
\end{footnotesize}}

\end{document}